\documentclass[twocolumn,showpacs,amsmath,amssymb]{revtex4}
\usepackage{appendix}
\usepackage{graphicx}
\usepackage{dcolumn}
\usepackage{bm}
\def\b{\begin{equation}}
\def\e{\end{equation}}
\begin{document}
\title{Electromagnetic Energy Density in Hyperbolic Metamaterials}

\author{Afshin Moradi$^{1}$}
\email{a.moradi@kut.ac.ir} 
\author{Pi-Gang Luan$^{2}$}
\email{pgluan@dop.ncu.edu.tw} 
\affiliation{%
\emph{$^{1}$Department of Engineering Physics, Kermanshah University of Technology, Kermanshah, Iran\\ $^{2}$Department of Optics and Photonics, National Central University, Jhongli District, Taoyuan City 320, Taiwan  }
}%

\begin{abstract}
We derive the energy density associated with an electromagnetic wave passing through a hyperbolic metamaterial (HMM). Both types of HMMs are studied. By considering a dispersive and absorbing HMM as an effective uniaxial crystal, we find that under the influence of an electromagnetic wave, the responses of type I (type II) HMM in the directions perpendicular to and parallel to the optical axis are similar to those of Lorentz (Drude) and Drude (Lorentz) media, respectively. Numerical examples are presented to reveal the general characteristics of the direction-dependent energy storage capacity of both types of HMMs.
\end{abstract}

 \maketitle
\section{Introduction}
\label{sec.1}

Metamaterials are artificial media that usually refers to arrays of wires and split-ring resonators (SRRs) \cite{D.R.S4184} and have unusual optical properties such as negative refractive index \cite{J.B.P3966}, subwavelength imaging \cite{X.Z435} and indefinite permittivity \cite{D.R.S1032}. The calculations of electromagnetic energy density in such metamaterials are of interest \cite{T.J.C205106, P.G.L675, P.G.L075016, J.C163493, A.M634}. However, the contradictory results of stored electromagnetic energy density in wire-SRR metamaterials have been the subject of theoretical controversy \cite{R.R309, S.A.T231, P.M.T92, A.D.B165110}. Note that in the absence of damping effects, the electromagnetic energy density of such media can be obtained correctly by considering the adiabatically varying electromagnetic field \cite{L.D.L}. In the presence of damping effects in wire-SRR metamaterials, on the other hand, difficulties arise in attempting to calculate the energy associated with an electromagnetic wave passing through such systems, and different results by two different methods were obtained by several researchers \cite{R.R309, S.A.T231, P.M.T92, A.D.B165110}. The question then becomes, what is the correct result for the energy density in wire-SRR metamaterials? In this way, one of us derived the electromagnetic energy density formula consistent with the Landau formula in the mentioned media \cite{P.G.L046601}.

A hyperbolic metamaterial (HMM) refers to an extremely anisotropic uniaxial optical media and has received much attention recently \cite{C.R.S4229, A.P948, P.S14, E.N214, L.F1, O.T463001, I.I.S}. An HMM is an anisotropic medium that has opposite signs of the two principal values of the permittivity tensor along and perpendicular to the optic axis. Mathematically, the dispersion of an electromagnetic wave in such a medium has a hyperbolic shape. By definition, there are two types of HMMs. Type I with a predominantly dielectric nature, and type II with a predominantly metallic nature. In an appropriate frequency region, a periodic array of metallic nanowires embedded in a host dielectric matrix is an example of an HMM of type I, and a multilayer medium consisting of dielectric and metallic layers is an example of an HMM of type II.

Recently, one of us \cite{P.G.L863} studied the propagation of electromagnetic energy in the multilayer HMMs in a simple case, when metallic layers have a permittivity of standard Drude model form, i.e., the constant part is assumed to be one. The results show that a multilayer HMM (as an HMM of type II in the appropriate frequency region) behaves like Drude and Lorentz medium, respectively, when the electric field is parallel and perpendicular to the layers. However, the problem of electromagnetic energy density in a nanowire HMM (as an HMM of type I) have not yet been studied. This is the main motivation of our present study. In this way, we obtain the energy density associated with an electromagnetic wave passing through an HMM, using a simple approach. In particular, both types of HMM are considered. For benefit of the reader, we also show that the electromagnetic response of a composite of metallic nanospheres embedded in a host matrix is similar to that of a Lorentz medium (see Appendix).

\section{Basic equations}
\label{sec.2}
Let us consider a dispersive and absorbing HMM as an effective uniaxial crystal. In the presence of an oscillating electric field $\textbf{E}=\textbf{E}_{\parallel}+\textbf{E}_{\bot}$ of frequency $\omega$ and long-wavelength vibration, the motion of the effective lattice of this effective uniaxial crystal may be represented by the Lorentz equations \cite{R.L233}
\b \label{2.1} m\ddot{\textbf{r}}_{\parallel}+m\gamma\dot{\textbf{r}}_{\parallel}+m\omega_{\mathrm{0}\parallel}^{2}\textbf{r}_{\parallel}=q\textbf{E}_{\parallel}  \;,\e
\b \label{2.2} m\ddot{\textbf{r}}_{\bot}+m\gamma\dot{\textbf{r}}_{\bot}+m\omega_{\mathrm{0}\bot}^{2}\textbf{r}_{\bot}=q\textbf{E}_{\bot}  \;,\e
where $q$ is the electric charge, $m$ is the effective mass of each electric charge, $\textbf{r}=\textbf{r}_{\parallel}+\textbf{r}_{\bot}$ is the displacement of the oscillators, $\omega_{\mathrm{0}\bot}$ ($\omega_{\mathrm{0}\parallel}$) is the resonance frequency of the charges and $\gamma$
is the damping frequency. Note $E_{x}=E_{y}=E_{\bot}$ and $E_{z}=E_{\parallel}$ and the $z$-axis is along the optical axis of the effective uniaxial crystal.

Suppose that there are $N$ oscillators in the effective uniaxial crystal volume $V$ and let us represent the effect of the high resonances by a real constant background dielectric constants $\varepsilon_{\infty\parallel}$ and $\varepsilon_{\infty\bot}$. Then, we have
\b \label{2.3} \textbf{P}_{\parallel}=\dfrac{Nq}{V}\textbf{r}_{\parallel}+\varepsilon_{0}\left( \varepsilon_{\infty\parallel}-1\right)\textbf{E}_{\parallel}   \;, \e
\b \label{2.4} \textbf{P}_{\bot}=\dfrac{Nq}{V}\textbf{r}_{\bot}+\varepsilon_{0}\left( \varepsilon_{\infty\bot}-1\right)\textbf{E}_{\bot}   \;, \e
and the components of the relative permittivity tensor determined by Eqs. \eqref{2.1}-\eqref{2.4} are
\b \label{2.5}\varepsilon_{\parallel}=1+\dfrac{P_{\parallel}}{\varepsilon_{0}E_{\parallel}} =\varepsilon_{\infty\parallel}\left( 1-\dfrac{F_{\parallel}\omega_{0\parallel}^{2}}{\omega(\omega+i\gamma)-\omega_{0\parallel}^{2}}\right) \;, \e
\b \label{2.6} \varepsilon_{\bot}=1+\dfrac{P_{\bot}}{\varepsilon_{0}E_{\bot}} =\varepsilon_{\infty\bot}\left( 1-\dfrac{F_{\bot}\omega_{0\bot}^{2}}{\omega(\omega+i\gamma)-\omega_{0\bot}^{2}}\right)    \;, \e
where $ F_{\parallel}=\omega_{\mathrm{p}\parallel}^{2}/\omega_{0\parallel}^{2}$ ($ F_{\bot}=\omega_{\mathrm{p}\bot}^{2}/\omega_{0\bot}^{2}$) measures the strength of the effective lattice resonance and $ \omega_{\mathrm{p}\parallel}=\left(Nq^{2}/\varepsilon_{0}\varepsilon_{\infty\parallel}mV \right)^{1/2} $ and $ \omega_{\mathrm{p}\bot}=\left(Nq^{2}/\varepsilon_{0}\varepsilon_{\infty\bot}mV \right)^{1/2} $. Using the derived formula by Loudon \cite{R.L233} for the energy and damping characteristics of the bulk electromagnetic waves in the bulk crystals, the time-averaged total energy density associated
with a wave can be written
\b \label{2.7} U=U_{\mathrm{EM}}+U_{\mathrm{K}}+U_{\mathrm{P}}\;,\e
where
\b \label{2.8} U_{\mathrm{EM}}=\dfrac{1}{4}\left[\varepsilon_{0}\varepsilon_{\infty\parallel}\vert \textbf{E}_{\parallel}\vert^{2}+\varepsilon_{0}\varepsilon_{\infty\bot}\vert \textbf{E}_{\bot}\vert^{2}+\mu_{0} \vert \textbf{H}\vert^{2}\right] \;,\e
is the electromagnetic energy density
\b \label{2.9} U_{\mathrm{K}}=\dfrac{1}{4}\dfrac{Nm}{V}\left(\vert \dot{\textbf{r}}_{\parallel}\vert^{2}+\vert \dot{\textbf{r}}_{\bot}\vert^{2} \right) \;,\e
is the kinetic energy density of the effective uniaxial crystal, and
\b \label{2.10} U_{\mathrm{P}}=\dfrac{1}{4}\dfrac{Nm}{V}\left( \omega_{0\parallel}^{2}\vert \textbf{r}_{\parallel}\vert^{2}+\omega_{0\bot}^{2}\vert \textbf{r}_{\bot}\vert^{2}\right) \;.\e
is the potential energy density of the effective uniaxial crystal. Therefore, in general the total energy includes two parts: the energies from the electric and magnetic fields themselves, and the energies from the medium response, i.e., the kinetic and potential energies of the charges present under the influence of the electromagnetic wave \cite{X.Z1853}. Also, the time-averaged power loss density can be written as
\begin{eqnarray} \label{2.11} P_{\mathrm{loss}}&=&\dfrac{Nm\gamma}{2V}\left(\vert \dot{\textbf{r}}_{\parallel}\vert^{2}+ \vert \dot{\textbf{r}}_{\bot}\vert^{2}\right)\nonumber \\ &=&2\gamma U_{\mathrm{K}}  \;.\end{eqnarray}
Using Eqs. \eqref{2.1} and \eqref{2.2} to eliminate $\textbf{r}_{\parallel}$ and $\textbf{r}_{\bot}$ from Eqs. \eqref{2.7} and \eqref{2.11}, we obtain
\b \label{2.12} U=\dfrac{1}{4}\left[\Xi_{\parallel}   \vert \textbf{E}_{\parallel}\vert^{2}+\Xi_{\bot} \vert \textbf{E}_{\bot}\vert^{2}+\mu_{0} \vert \textbf{H}\vert^{2}\right] \;,\e
where
\begin{equation*}\Xi_{\parallel}=\varepsilon_{0}\varepsilon_{\infty\parallel}\left( 1+\dfrac{F_{\parallel}\omega_{0\parallel}^{2}\left( \omega^{2}+\omega_{0\parallel}^{2}\right)}{\left( \omega^{2}-\omega_{0\parallel}^{2}\right) ^{2} +\gamma^{2}\omega^{2}} \right)\;,\end{equation*}
\begin{equation*}\Xi_{\bot}=\varepsilon_{0}\varepsilon_{\infty\bot}\left(1+\dfrac{F_{\bot}\omega_{0\bot}^{2}\left( \omega^{2}+\omega_{0\bot}^{2}\right)}{\left( \omega^{2}-\omega_{0\bot}^{2}\right)^{2} +\gamma^{2}\omega^{2}} \right)\;,\end{equation*}
and
\begin{eqnarray} \label{2.13} P_{\mathrm{loss}}&=&\dfrac{\gamma\varepsilon_{0}}{2}\left[\Gamma_{\parallel} \vert \textbf{E}_{\parallel}\vert^{2}+\Gamma_{\bot} \vert \textbf{E}_{\bot}\vert^{2}\right]\nonumber \\ &=&\dfrac{\omega\varepsilon_{0}}{2}\operatorname{Im}\left[\varepsilon_{\parallel} \vert \textbf{E}_{\parallel}\vert^{2}+\varepsilon_{\bot} \vert \textbf{E}_{\bot}\vert^{2}\right]\;, \end{eqnarray}
where
\begin{equation*}\Gamma_{\parallel}=\varepsilon_{\infty\parallel}\dfrac{F_{\parallel}\omega_{0\parallel}^{2}\omega^{2}}{\left( \omega^{2}-\omega_{0\parallel}^{2}\right)^{2} +\gamma^{2}\omega^{2}}\;, \end{equation*}
\begin{equation*}\Gamma_{\bot}=\varepsilon_{\infty\bot}\dfrac{F_{\bot}\omega_{0\bot}^{2}\omega^{2}}{\left( \omega^{2}-\omega_{0\bot}^{2}\right)^{2} +\gamma^{2}\omega^{2}}\;. \end{equation*}If the losses are negligible, the time-averaged electromagnetic energy density, for a monochromatic (single frequency) electromagnetic field, is \cite{L.D.L}
\b \label{2.14} U=\dfrac{\varepsilon_{0}}{4}\left[ \dfrac{d (\omega \varepsilon_{\parallel}) }{d\omega} \vert \textbf{E}_{\parallel}\vert^{2}+\dfrac{d (\omega \varepsilon_{\bot}) }{d\omega} \vert \textbf{E}_{\bot}\vert^{2}\right] +\dfrac{\mu_{0}}{4} \vert \textbf{H}\vert^{2}\;.\e
Setting $\gamma=0$ in Eqs. \eqref{2.5} and \eqref{2.6}, and using Eq.
\eqref{2.14}, leads to
\begin{multline} \label{2.15} U=\dfrac{1}{4}\varepsilon_{0}\varepsilon_{\infty\parallel}\left(1+\dfrac{F_{\parallel}\omega_{0\parallel}^{2}\left( \omega^{2}+\omega_{0\parallel}^{2}\right)}{\left( \omega^{2}-\omega_{0\parallel}^{2}\right)^{2} } \right)  \vert \textbf{E}_{\parallel}\vert^{2}\\+\dfrac{1}{4}\varepsilon_{0}\varepsilon_{\infty\bot}\left(1+\dfrac{F_{\bot}\omega_{0\bot}^{2}\left( \omega^{2}+\omega_{0\bot}^{2}\right)}{\left( \omega^{2}-\omega_{0\bot}^{2}\right)^{2} } \right) \vert \textbf{E}_{\bot}\vert^{2}\\+\dfrac{1}{4}\mu_{0} \vert \textbf{H}\vert^{2}\;,\end{multline}
that is equal to the result as that obtained by setting $\gamma=0$ in Eq. \eqref{2.12}.

\section{Nanowire hyperbolic metamaterials}
\label{sec.3}
Consider a periodic array of metallic nanowires with axes parallel to
$z$-axis embedded in a host dielectric matrix with the dielectric constant $\varepsilon_{\mathrm{d}}$. The $z$-axis is along the optical axis. Let $f$ be the filling fraction of the metallic nanowires in a unit cell satisfying $0<f<1$. This periodic array can be used to construct an electric HMM of type I (in an
appropriate frequency region), with effective permittivity given by
\begin{eqnarray} \label{3.1} \varepsilon_{\mathrm{\parallel}}&=&f\varepsilon_{\mathrm{m}}+(1-f)\varepsilon_{\mathrm{d}}\;,\\
\label{3.2} \varepsilon_{\bot}&=&\varepsilon_{\mathrm{d}}\dfrac{(1+f)\varepsilon_{\mathrm{m}}+(1-f)\varepsilon_{\mathrm{d}}}{(1-f)\varepsilon_{\mathrm{m}}+(1+f)\varepsilon_{\mathrm{d}}}\;.\end{eqnarray}
where
\b \label{3.3}\varepsilon_{\mathrm{m}}=\varepsilon_{\infty}\left( 1-\dfrac{\omega_{\mathrm{p}0}^{2}}{\omega\left(\omega+i\gamma \right)}\right) \;,\e 
shows the relative permittivity of a metallic nanowire with $\varepsilon_{\infty}$ that is the permittivity in
high-frequency and $\omega_{\mathrm{p}0}$ that is the plasma frequency. Here we have relaxed the restrictions in the previous investigation \cite{P.G.L863} by considering $\varepsilon_{\infty}$ in Eq. \eqref{3.3}, therefore the formulas might provide more realistic applications.

Now, we can rewrite Eqs. \eqref{3.1} and \eqref{3.2} as
Eqs. \eqref{2.5} and \eqref{2.6}, respectively, where \begin{eqnarray}
 \label{3.4}\varepsilon_{\infty\parallel}&=&f\varepsilon_{\infty}+(1-f)\varepsilon_{\mathrm{d}}\;,\\
\label{3.5}\varepsilon_{\infty\bot}&=&\varepsilon_{\mathrm{d}}\dfrac{(1+f)\varepsilon_{\infty}+(1-f)\varepsilon_{\mathrm{d}}}{(1-f)\varepsilon_{\infty}+(1+f)\varepsilon_{\mathrm{d}}}\;,\\\label{3.6}F_{\bot}&=&\dfrac{4f\varepsilon_{\mathrm{d}}}{(1-f)\left[ (1+f)\varepsilon_{\infty}+(1-f)\varepsilon_{\mathrm{d}}\right] }\;,\\
 \label{3.7}\omega_{\mathrm{p}\parallel}^{2}&=&\dfrac{f\varepsilon_{\infty}\omega_{\mathrm{p}0}^{2}}{f\varepsilon_{\infty}+(1-f)\varepsilon_{\mathrm{d}}}\;,\\
 \label{3.8}\omega_{0\bot}^{2}&=&\dfrac{(1-f)\varepsilon_{\infty}\omega_{\mathrm{p}0}^{2}}{(1-f)\varepsilon_{\infty}+(1+f)\varepsilon_{\mathrm{d}}}\;,\\
 \label{3.9}\omega_{0\parallel}&=&0\;. \end{eqnarray}
These equations indicate that the responses of a nanowire HMM in the directions perpendicular to and parallel to the optical axis are similar to those of Lorentz and Drude media, respectively. Using the above relations, we can obtain energy relations of HMMs of type I according to the Eqs. \eqref{2.7}-\eqref{2.15} of the previous section. For the case $f\ll1$, nanowires are well separated by large distances and do not feel
each other by electromagnetic interactions. Then from Eqs. \eqref{3.4}-\eqref{3.8} we find:
\begin{equation*} \varepsilon_{\infty\parallel}=\varepsilon_{\mathrm{d}}  \;,\;\;\;\;\varepsilon_{\infty\bot}=\varepsilon_{\mathrm{d}} \;,\end{equation*} \b \label{3.10} \omega_{\mathrm{p}\parallel}^{2}=\dfrac{f\varepsilon_{\infty}\omega_{\mathrm{p}0}^{2}}{\varepsilon_{\mathrm{d}}}\;,\;\;\;\;\; \omega_{0\bot}^{2}=\dfrac{\varepsilon_{\infty}\omega_{\mathrm{p}0}^{2}}{\varepsilon_{\infty}+\varepsilon_{\mathrm{d}}}\;,\;\;\;\;\;F_{\bot}=\dfrac{4f\varepsilon_{\mathrm{d}}}{ \varepsilon_{\infty}+\varepsilon_{\mathrm{d}}}\;, \e
where $\varepsilon_{\infty}\omega_{\mathrm{p}0}/\sqrt{\varepsilon_{\infty}+\varepsilon_{\mathrm{d}}}$ is the frequency of dipolar resonance of a single metallic nanowire surrounded by a dielectric medium \cite{A.M064502}.

Note that for an electric HMM of type I we should have ${\rm Re}\left[ \varepsilon_{\mathrm{\parallel}}\right] <0$ and ${\rm Re}\left[ \varepsilon_{\mathrm{\bot}}\right] >0$ and the imaginary parts of them must be small enough to be negligible. Therefore, using Eqs. \eqref{2.5} and \eqref{2.6} we find ${\rm Re}\left[ \varepsilon_{\mathrm{\parallel}}\right] <0$, if $\omega<\sqrt{\omega_{\mathrm{p}\parallel}^{2}-\gamma^{2}}$, and ${\rm Re}\left[ \varepsilon_{\mathrm{\bot}}\right] >0$, if $\omega<\omega_{0\bot}$, that means 
\b \label{3.11}\omega<\mathrm{min}\left( \sqrt{\omega_{\mathrm{p}\parallel}^{2}-\gamma^{2}},\omega_{0\bot}\right) \;.\e

\section{Multilayer hyperbolic metamaterials}
\label{sec.4}
Consider a multilayer structure consisting of isotropic metal/dielectric layers. The $z$-axis is directed along the optical axis. Let $\varepsilon_{\mathrm{m}}$ ($\varepsilon_{\mathrm{d}}$) be the relative permittivity of the metal (dielectric) layer, and let $f$ be the filling ratio of the metal layer satisfying $0<f<1$. Also, Eq. \eqref{3.3} shows the relative permittivity of a metallic layer. This multilayer structure can be used to construct an electric HMM of type II (in an appropriate frequency region), with effective permittivity given by
\begin{eqnarray} \label{4.1} \varepsilon_{\mathrm{\parallel}}&=&\dfrac{\varepsilon_{\mathrm{m}}\varepsilon_{\mathrm{d}}}{(1-f)\varepsilon_{\mathrm{m}}+f\varepsilon_{\mathrm{d}}}\;,\\
\label{4.2} \varepsilon_{\mathrm{\bot}}&=&f\varepsilon_{\mathrm{m}}+(1-f)\varepsilon_{\mathrm{d}}\;.\end{eqnarray}
We can rewrite Eqs. \eqref{4.1} and \eqref{4.2} as
Eqs. \eqref{2.5} and \eqref{2.6}, respectively, where \cite{P.G.L863} 
\begin{eqnarray} \label{4.3}\varepsilon_{\infty\bot}&=&f\varepsilon_{\infty}+(1-f)\varepsilon_{\mathrm{d}}\;,\\
 \label{4.4}\varepsilon_{\infty\parallel}&=&\dfrac{\varepsilon_{\infty}\varepsilon_{\mathrm{d}}}{(1-f)\varepsilon_{\infty}+f\varepsilon_{\mathrm{d}}}\;,\\
 \label{4.5}F_{\parallel}&=&\dfrac{f}{1-f}\dfrac{\varepsilon_{\mathrm{d}}}{\varepsilon_{\infty}}\;,\\
 \label{4.6}\omega_{\mathrm{p}\bot}^{2}&=&\dfrac{f\varepsilon_{\infty}\omega_{\mathrm{p}0}^{2}}{f\varepsilon_{\infty}+(1-f)\varepsilon_{\mathrm{d}}}\;,\\
 \label{4.7}\omega_{0\parallel}^{2}&=&\dfrac{(1-f)\varepsilon_{\infty}\omega_{\mathrm{p}0}^{2}}{(1-f)\varepsilon_{\infty}+f\varepsilon_{\mathrm{d}}}\;,\\
 \label{4.8}\omega_{0\bot}&=&0\;.\end{eqnarray}
These equations indicate that the responses of a multilayer HMM in the directions perpendicular to and parallel to the optical axis are similar to those of Drude and Lorentz media, respectively. Using the above relations, we can obtain energy relations of HMMs of type II according to the Eqs. \eqref{2.7}-\eqref{2.15} of Sec. \ref{sec.2}. Again, if we consider $f\ll1$ then from Eqs. \eqref{4.3}-\eqref{4.7} we find:
\begin{equation*} \varepsilon_{\infty\bot}=\varepsilon_{\mathrm{d}}  \;,\;\;\;\;\varepsilon_{\infty\parallel}=\varepsilon_{\mathrm{d}} \;,\end{equation*} \b \label{4.9} \omega_{\mathrm{p}\bot}^{2}=\dfrac{f\varepsilon_{\infty}\omega_{\mathrm{p}0}^{2}}{\varepsilon_{\mathrm{d}}}\;,\;\;\;\;\;\omega_{0\parallel}^{2}=\omega_{\mathrm{p}0}^{2}\;,\;\;\;\;\;F_{\parallel}=f\dfrac{\varepsilon_{\mathrm{d}}}{\varepsilon_{\infty}}\;, \e
where $\omega_{\mathrm{p}0}$ is the resonance frequency of a single metallic layer in a dielectric medium. This indicates that the resonance effect also happens in the metallic layers of the system, however as a result we note that this resonance frequency is free from the effect of permittivity of the dielectric medium.
\begin{figure}[!htb]
\centering
\includegraphics[width=7cm,clip]{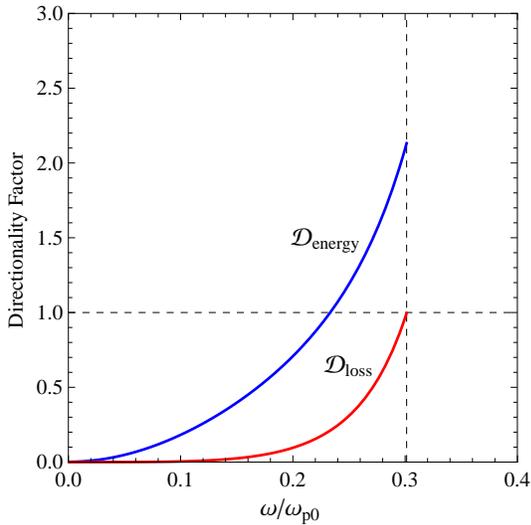}
\caption{Variation of $\mathcal{D}_{\mathrm{energy}}$ (blue curves), and $\mathcal{D}_{\mathrm{loss}}$ (red curves) with respect to the dimensionless frequency $\omega/\omega_{\mathrm{p}0}$ for a nanowire HMM as a type I HHM. Here, the filling fraction is $f=0.2$, the relative permittivity of the host dielectric matrix is $\varepsilon_{\mathrm{d}}=2.5$, $\varepsilon_{\infty}=1$, and $\gamma=0.01\omega_{\mathrm{p}0}$. Therefore, we have 
$\varepsilon_{\infty\parallel}=2.2$, $\varepsilon_{\infty\bot}=2.11$,
$\omega_{\mathrm{p}\parallel}=0.3\omega_{\mathrm{p}0}$,
$\omega_{0\bot}=0.459\omega_{\mathrm{p}0}$, and $F_{\bot}=0.78 $. Also, the vertical dashed line shows $\omega/\omega_{\mathrm{p}0}=0.301$.}
\label{fig.1}
\end{figure}
 \begin{figure}[!htb]
\centering
\includegraphics[width=7cm,clip]{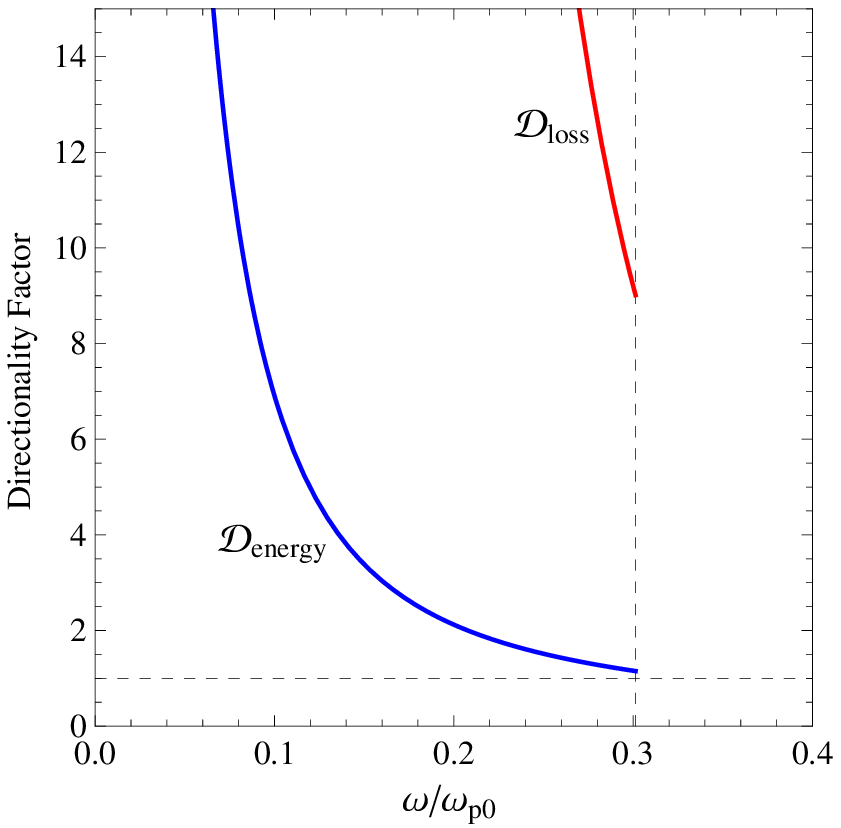}
\caption{Variation of $\mathcal{D}_{\mathrm{energy}}$ (blue curves), and $\mathcal{D}_{\mathrm{loss}}$ (red curves) with respect to the dimensionless frequency $\omega/\omega_{\mathrm{p}0}$ for a multilayer HMM as a type II HHM. Here, the filling fraction is $f=0.2$, the relative permittivity of the dielectric layers is $\varepsilon_{\mathrm{d}}=2.5$, $\varepsilon_{\infty}=1$, and $\gamma=0.01\omega_{\mathrm{p}0}$. Therefore, we have $\varepsilon_{\infty\bot}=2.2$, $\varepsilon_{\infty\parallel}=1.92$, $\omega_{\mathrm{p}\bot}=0.3\omega_{\mathrm{p}0}$, $\omega_{0\parallel}=0.78\omega_{\mathrm{p}0}$, and $F_{\parallel}=0.625$. Also, the vertical dashed line shows $\omega/\omega_{\mathrm{p}0}=0.301$.}
\label{fig.2}
\end{figure}

Note that for an electric HMM of type II we should have ${\rm Re}\left[\varepsilon_{\mathrm{\parallel}}\right] >0$ and ${\rm Re}\left[\varepsilon_{\mathrm{\bot}}\right] <0$. Therefore, using Eqs. \eqref{2.5} and \eqref{2.6} we find ${\rm Re}\left[\varepsilon_{\parallel}\right] >0$, if $\omega<\omega_{0\parallel}$, and
${\rm Re}\left[ \varepsilon_{\bot}\right] <0$, if $\omega<\sqrt{\omega_{\mathrm{p}\bot}^{2}-\gamma^{2}}$, that means 
\b \label{4.10}\omega<\mathrm{min}\left( \sqrt{\omega_{\mathrm{p}\bot}^{2}-\gamma^{2}},\omega_{0\parallel}\right) \;.\e

\section{Numerical results}
\label{sec.5}
In this section, numerical results for the distribution of electromagnetic energy density in an HMM are presented. We assume that the filling fraction is $f=0.2$, the relative permittivity of the dielectric medium is $\varepsilon_{\mathrm{d}}=2.5$, $\varepsilon_{\infty}=1$, and $\gamma=0.01\omega_{\mathrm{p}0}$.
Therefore, we use the following values for the parameters appearing in Eqs. \eqref{2.12} and \eqref{2.13}:
$\varepsilon_{\infty\parallel}=2.2$, $\varepsilon_{\infty\bot}=2.11$,
$\omega_{\mathrm{p}\parallel}=0.3\omega_{\mathrm{p}0}$,
$\omega_{0\bot}=0.459\omega_{\mathrm{p}0}$, and $F_{\bot}=0.78 $ for a periodic array of metallic nanowires as a type I HMM, and 
$\varepsilon_{\infty\bot}=2.2$, $\varepsilon_{\infty\parallel}=1.92$ $\omega_{\mathrm{p}\bot}=0.3\omega_{\mathrm{p}0}$, $\omega_{0\parallel}=0.78\omega_{\mathrm{p}0}$, and $F_{\parallel}=0.625$ for a multilayer HMM as a type II HMM. 

To study the corresponding dispersion/anisotropy and absorption effects in the energy density and power loss, we use the directionality factors \cite{P.G.L863}, as 
$ \mathcal{D}_{\mathrm{energy}}=\Xi_{\bot}/\Xi_{\parallel}$,
where $\Xi_{\parallel}$ and $\Xi_{\bot}$ are the coefficients $\vert \textbf{E}_{\parallel}\vert^{2}$ and $\vert \textbf{E}_{\bot}\vert^{2}$ in Eq. \eqref{2.12}, respectively, and
$\mathcal{D}_{\mathrm{loss}}=\Gamma_{\bot}/\Gamma_{\parallel}$,
where $\Gamma_{\parallel}$ and $\Gamma_{\bot}$ are the coefficients $\vert \textbf{E}_{\parallel}\vert^{2}$ and $\vert \textbf{E}_{\bot}\vert^{2}$ in Eq. \eqref{2.13}, respectively.

Figs. \ref{fig.1} and \ref{fig.2} show the variation of $\mathcal{D}_{\mathrm{energy}}$ (blue curves), and $\mathcal{D}_{\mathrm{loss}}$ (red curves) with respect to the dimensionless frequency $\omega/\omega_{\mathrm{p}0}$ for HHMs of types I and II, respectively. From Fig. \ref{fig.1} one can see the absorption property of type I HMM becomes
isotropic with respect to the electric field at \begin{equation*}
\dfrac{\omega}{\omega_{\mathrm{p}0}}=\sqrt{\dfrac{f^{2}}{\left[ f+(1-f)\varepsilon_{\mathrm{d}}\right]^{2} }-\gamma^{2}}=0.301\;,\end{equation*} while the energy storage ability of the medium becomes 
isotropic with respect to the electric field for $\omega/\omega_{\mathrm{p}0}<0.301$. Also, $\mathcal{D}_{\mathrm{loss}}$ and $\mathcal{D}_{\mathrm{energy}}$ decrease with decreasing the value of $\omega/\omega_{\mathrm{p}0}$. Furthermore, from Fig. \ref{fig.2} it is clear that the energy storage ability of the type II HMM becomes almost
isotropic with respect to the electric field at $\omega/\omega_{\mathrm{p}0}=0.301$. Below this frequency, the type II HMM can store more energy if the electric field direction is parallel to the layers.

\section{Conclusion}
\label{sec.6}
Although it seems that the structure of an HMM is much simpler than the wire-SRR and chiral metamaterials, however, the effective permittivities of HMMs are more complex than those of wire-SRR and chiral metamaterials and investigation of electromagnetic energy density in such media may be difficult. To remedy this difficulty, we have obtained familiar forms for the effective permittivities of nanowire HMMs and multilayer HMMs, i.e., similar to the Lorentz and Drude media. These new forms for the effective permittivities of the HMMs simply show that HMMs have different dynamical properties in the direction parallel and perpendicular to their optical axis. Using these new forms of the effective permittivities of HMMs, we have derived the energy density associated with an electromagnetic wave passing through a HMM. In this way, HMM of types I and II have been treated, separately.

\renewcommand{\theequation}{A-\arabic{equation}} 
\setcounter{equation}{0}
\section*{Appendix: Lorentz model of a composite of metallic nanospheres }
 Consider a composite of Drude metallic nanospheres with relative dielectric constant $\varepsilon_{\mathrm{m}}=\varepsilon_{\infty}\left( 1-\omega_{\mathrm{p}0}^{2}/\omega\left(\omega+i\gamma \right)\right) $ embedded in a host matrix with relative dielectric
constant $\varepsilon_{\mathrm{d}}$. Let $f$ be the volume fraction of the embedded nanospheres satisfying $0<f<1$. For the relative effective permittivity of the present composite we have \cite{U.K, A.M}
\b \label{A1} \dfrac{\varepsilon_{\mathrm{eff}}-\varepsilon_{\mathrm{d}}}{\varepsilon_{\mathrm{eff}}+2\varepsilon_{\mathrm{d}}}=f\dfrac{\varepsilon_{\mathrm{m}}-\varepsilon_{\mathrm{d}}}{\varepsilon_{\mathrm{m}}+2\varepsilon_{\mathrm{d}}}\;,\e
or
\b \label{A2} \varepsilon_{\mathrm{eff}}=\varepsilon_{\mathrm{d}}\dfrac{\left( \varepsilon_{\mathrm{m}}+2 \varepsilon_{\mathrm{d}}\right)+2 f\left( \varepsilon_{\mathrm{m}}-\varepsilon_{\mathrm{d}}\right) }{\left( \varepsilon_{\mathrm{m}}+2 \varepsilon_{\mathrm{d}}\right)- f\left( \varepsilon_{\mathrm{m}}-\varepsilon_{\mathrm{d}}\right)}\;.\e
We can rewrite  Eq. \eqref{A2} as 
\b \label{A3}\varepsilon_{\mathrm{eff}}=\varepsilon_{\mathrm{b}}\left( 1-\dfrac{F\omega_{0}^{2}}{\omega(\omega+i\gamma)-\omega_{0}^{2}}\right) \;, \e
 where
\begin{eqnarray} \label{A4}\varepsilon_{\mathrm{b}}&=&\varepsilon_{\mathrm{d}}\dfrac{\left( \varepsilon_{\infty}+2 \varepsilon_{\mathrm{d}}\right)+2 f\left( \varepsilon_{\infty}-\varepsilon_{\mathrm{d}}\right) }{\left( \varepsilon_{\infty}+2 \varepsilon_{\mathrm{d}}\right)- f\left( \varepsilon_{\infty}-\varepsilon_{\mathrm{d}}\right)}\;,\\
 \label{A6}F&=&\dfrac{9f\varepsilon_{\mathrm{d}}}{(1-f)\left[ \left( \varepsilon_{\infty}+2 \varepsilon_{\mathrm{d}}\right)+2 f\left( \varepsilon_{\infty}-\varepsilon_{\mathrm{d}}\right) \right] }\;,\\
 \label{A5}\omega_{0}^{2}&=&\dfrac{(1-f)\varepsilon_{\infty}\omega_{\mathrm{p}0}^{2}}{\left( \varepsilon_{\infty}+2 \varepsilon_{\mathrm{d}}\right)- f\left( \varepsilon_{\infty}-\varepsilon_{\mathrm{d}}\right)}\;.\end{eqnarray}
For the case $f\ll1$ we find $\omega_{0}=\varepsilon_{\infty}\omega_{\mathrm{p}0}/\sqrt{\varepsilon_{\infty}+2\varepsilon_{\mathrm{d}}}$ that is the frequency of dipolar resonance of a single metallic nanosphere surrounded by a dielectric medium \cite{A.M849}.

\end{document}